%% file: NEOQEC_main.tex
\documentclass[prx,twocolumn,english,superscriptaddress,floatfix,nofootinbib]{revtex4-2}%,longbibliography
\usepackage[english]{babel}
\usepackage{graphicx}% Include figure files
\usepackage{dcolumn}% Align table columns on decimal point
\usepackage{bm}% bold math
\usepackage{siunitx}
\usepackage[super]{nth}
\usepackage{amsmath,amssymb,amsfonts}
\usepackage{textcomp}
\usepackage{xcolor}

\usepackage{tikz,braket}
\usepackage{lipsum}
\usepackage{diagbox}
\usepackage[RPvoltages]{circuitikz}
\usetikzlibrary{arrows,decorations.pathreplacing}

\usepackage{comment}
\usepackage{subfiles}
\usepackage{epsfig}
\usepackage{here}
\usepackage{multirow}
\usepackage{algorithm}
\usepackage{algorithmicx}
\usepackage{algpseudocode}
\usepackage{braket}
\usepackage{romannum}
\usepackage[stable]{footmisc}

\begin{document}

\preprint{APS/123-QED}

\title{NEO-QEC: Neural Network Enhanced Online\\Superconducting Decoder for Surface Codes}% Force line breaks with \\
\author{Yosuke Ueno}
 \email{ueno@hal.ipc.i.u-tokyo.ac.jp}
\affiliation{The University of Tokyo}%
\author{Masaaki Kondo}%
\affiliation{Keio University
}
\author{Masamitsu Tanaka}%
\affiliation{Nagoya University
}
\author{Yasunari Suzuki}%
\affiliation{NTT Computer and Data Science Laboratories
}
\author{Yutaka Tabuchi}%
\affiliation{RIKEN Center for Quantum Computing
}


\begin{abstract}

Quantum error correction (QEC) is essential for quantum computing to mitigate the effect of errors on qubits, and \textit{surface code} (SC) is one of the most promising QEC methods. 
Decoding SCs is the most computational expensive task in the control device of quantum computers (QCs), and many works focus on accurate decoding algorithms for SCs, including ones with neural networks (NNs).
Practical QCs also require low-latency decoding because slow decoding leads to the accumulation of errors on qubits, resulting in logical failures. 
For QCs with superconducting qubits, a practical decoder must be very power-efficient in addition to having high accuracy and low latency. In order to reduce the hardware complexity of QC, we are supposed to decode SCs in a cryogenic environment with a limited power budget, where superconducting qubits operate.

In this paper, we propose an NN-based accurate, fast, and low-power decoder capable of decoding SCs and lattice surgery (LS) operations with measurement errors on ancillary qubits. 
To achieve both accuracy and hardware efficiency of the SC decoder, we apply a binarized NN.
We design a neural processing unit (NPU) for the decoder with \textit{SFQ}-based digital circuits and evaluate it with a SPICE-level simulation. 
We evaluate the decoder performance by a quantum error simulator for the single logical qubit protection and the minimum operation of LS with code distances up to 13, and it achieves 2.5\% and 1.0\% accuracy thresholds, respectively.

\end{abstract}

\maketitle

%\tableofcontents

\subfile{./NEOQECchap/00intro}

\subfile{./NEOQECchap/01background}

\subfile{./NEOQECchap/02related_work}

\subfile{./NEOQECchap/03proposed}

%\subfile{./NEOQECchap/04implementation}
\subfile{./NEOQECchap/04implementation_counter}

\subfile{./NEOQECchap/05evaluation}

\subfile{./NEOQECchap/06conclusion}

%\nocite{*}
%\bibliographystyle{acm}
\bibliography{cite}% Produces the bibliography via BibTeX.

\end{document}

%% file: NEOQECchap/00intro.tex
\section{Introduction}
\label{sec:BNN_intro}

Quantum computers (QCs) are becoming an attractive computing paradigm as the number of implementable qubits increases because of their potential to solve meaningful quantum algorithms. One of the most important challenges to practical quantum computation is the fragility of qubits. As quantum error correction (QEC) is the unique method to reduce the effective error rate of quantum gates in a scalable manner, it has been extensively studied so far~\cite{Shor_physicalreview_1995,PhysRevLett.77.793,KITAEV20032,bravyi1998quantum,fowler2012surface}. QEC codes encode fault-tolerant logical qubits by using a set of multiple faulty physical qubits.

Surface code (SC)~\cite{bravyi1998quantum} is a promising candidate in practical QEC coding schemes. SC is implemented on a square grid of qubits and only requires interactions between geometrically adjacent physical qubits. These features simplify the hardware implementation and provide extendibility and high reliability in QEC. The decoding procedure is known to be a non-trivial task and requires a large amount of computational cost in classical computers.

In order to achieve a practical fault-tolerant quantum computation (FTQC), we need an SC decoder that satisfies the following three requirements: accuracy, latency, and scalability. The accuracy requirement indicates that the decoder must estimate both errors on physical qubits and measurement processes with high probability. The latency requirement indicates that the decoder should correct errors within a QEC cycle to prevent the accumulation of errors. The scalability requirement indicates that the decoder must support hundreds of logical qubits with interactions between them\cite{horsman2012surface,fowler2012surface}. Thus, the QEC coding/decoding schemes should be optimized in algorithmic and implementation perspectives to meet all requirements.

When targeting superconducting QCs, practical decoders must also satisfy a power requirement. A superconducting quantum circuit is one of the most promising QC implementations~\cite{arute2019quantum,Gong948}, and state-of-the-art superconducting QCs have over 100 qubits. They are operated in a cryogenic environment with an effective temperature of around ten milliKelvins to eliminate thermal noises in the device. While we place the qubits at the milliKelvin stage of a dilution refrigerator, the associated control electronics, including a QEC decoder, are supposed to be located at a higher temperature stage or outside the cryostat~\cite{cryogenic_quantum}. The spatial distance between the components demands many cables between different temperature stages, leading to the hardware complexity in wiring and latency in QEC. This point hinders scaling up QCs.
Although the QEC processing unit right next to the qubit chip alleviates this problem~\cite{2017micro_qureshi}, it is usually unrealistic under the restricted power budget in the lower temperature stages of a cryostat (\textit{e.g.}, tens of $\mu$W or around 1~W in the millikelvin stage or the 4-K stage, respectively). Therefore, an extraordinary low-power decoder is necessary. %The same problem applies to other types of solid-state qubits that operate in a cryogenic environment, \textit{e.g.} spin qubits in donor silicon and MOS-type double quantum dots.

In the decoding process of SCs, many works have proposed accurate decoding algorithms such as minimum weight perfect matching (MWPM)\cite{fowler2015minimum}, union find (UF)\cite{delfosee2017almost}, and renormalization group (RG)\cite{duclos2010fast}. These software-based decoders guarantee a polynomial-time solution; however, they are slow and not hardware efficient due to their high computational cost and do not meet the latency and power requirements. 
Several fast and low-cost decoding algorithms and their hardware-efficient implementations~\cite{torlai2017neural,delfosee2017almost,holmes2020nisq,ueno2021qecool,das2022afs,ueno2022qulatis} have been proposed to reduce the computational burden. In particular, the previous works based on a greedy matching algorithm and high-speed and low-power superconducting digital circuits~\cite{holmes2020nisq,ueno2021qecool,ueno2022qulatis} operate in a cryogenic environment to alleviate wires between superconducting qubits and satisfy the latency and power requirements. Among them, the QULATIS decoder~\cite{ueno2022qulatis} also meets the scalability requirement because it supports lattice surgery (LS)\cite{horsman2012surface}. However, while these decoders have the desirable properties previously described, their decoding accuracy is lower than that of software decoders due to their greedy nature.

For the last several years, researchers have studied neural network (NN) techniques for decoding SCs\cite{varsamopoulos2017decoding,varsamopoulos2019comparing,davaasuren2020general,varsamopoulos2020decoding,meinerz2021scalable,bhoumik2021efficient,gicev2021scalable}. Generally, the performance of NN-based decoders is expected to be superior to that of conventional software decoders, such as MWPM or UF because they can incorporate the correlation between $X$- and $Z$-errors into decoding SCs. In addition, several works proposed methods of constructing an accurate and scalable decoder by combining NNs with other decoders\cite{meinerz2021scalable,bhoumik2021efficient,gicev2021scalable}.
However, all previous NN-based decoders do not meet the latency and power requirements.

In this paper, we propose a new neural network based decoding scheme called Neural network Enhanced Online Quantum Error Correction (NEO-QEC) that meets the four requirements required for superconducting QCs. We extend the previous NN decoder based on convolutional neural networks (CNNs)~\cite{gicev2021scalable} and combine it with the superconducting-digital-circuits-based decoders\cite{ueno2021qecool,ueno2022qulatis}. We aim to design a decoder based on the algorithm to achieve the following features simultaneously:
\begin{enumerate}
    \item Improved error correction performance.
    \item Low latency to prevent the accumulation of errors.
    \item Capability for decoding three-dimensional (3-D) complex SC lattices to handle LS with measurement errors.
    \item Power efficiency suitable for operation in a cryogenic environment.
\end{enumerate}
There are many problems in satisfying the latency and power requirements for NN-based decoders because NNs generally require heavy computational resources.
To design a fast and low-power NN decoder, we use quantization \cite{Gholami2021survey} or binarization\cite{Rastegari2016XNORNetIC} techniques.

The contributions of this paper are summarized as follows:
\begin{itemize}
  \item We propose the NEO-QEC decoding algorithm that is capable of LS with measurement errors by combines NNs and existing decoders.
  \item We use a binarization technique to implement a lightweight NN decoder with moderate accuracy degradation.
  \item We evaluate the error correction performance of the NEO-QEC by a quantum error simulator.
  \item We design an ultra-low-power and fast neural processing unit (NPU) with superconducting digital circuits and evaluate its performance with a SPICE-level simulator.
\end{itemize}

The remainder of the paper is organized as follows: we begin with a background of this paper in Section~\ref{sec:BNN_background}.
Then, we show an overview of prior NN-based decoders and superconducting-digital-circuits-based decoders in Section~\ref{sec:related_work}. In Section~\ref{sec:BNN_proposed} we describe our new decoding algorithm. In Section~\ref{sec:BNN_hardware_section} we show the superconducting circuit design of the NPU. We then show the evaluation results of our decoder performance in Section~\ref{sec:BNN_evaluation} and conclude the paper with future works in Section~\ref{sec:BNN_conclusion}.

%% file: NEOQECchap/01background.tex
\section{Background}
\label{sec:BNN_background}

\subfile{./bg_surface_code}

\subfile{./bg_lattice_surgery}

\subfile{./bg_SFQ}

%% file: NEOQECchap/bg_surface_code.tex
\subsection{Surface code\label{sec:bg_surface_code}
}

Surface code (SC)~\cite{bravyi1998quantum,kitaev1997quantum} is one of the most promising QEC codes; SC can reduce logical errors effectively and be implemented simply with physical operations on geometrically adjacent qubits located on a two-dimensional (2-D) grid. SC consists of two types of physical qubits: data and ancillary. Data qubits are used to represent a logical qubit. Ancillary qubits are utilized to check the parity of errors on the neighboring data qubits. This parity check is called a stabilizer measurement, and its binary outcome is called a syndrome value. 

\begin{figure}[tb]
    \centering
    \includegraphics[width=\linewidth]{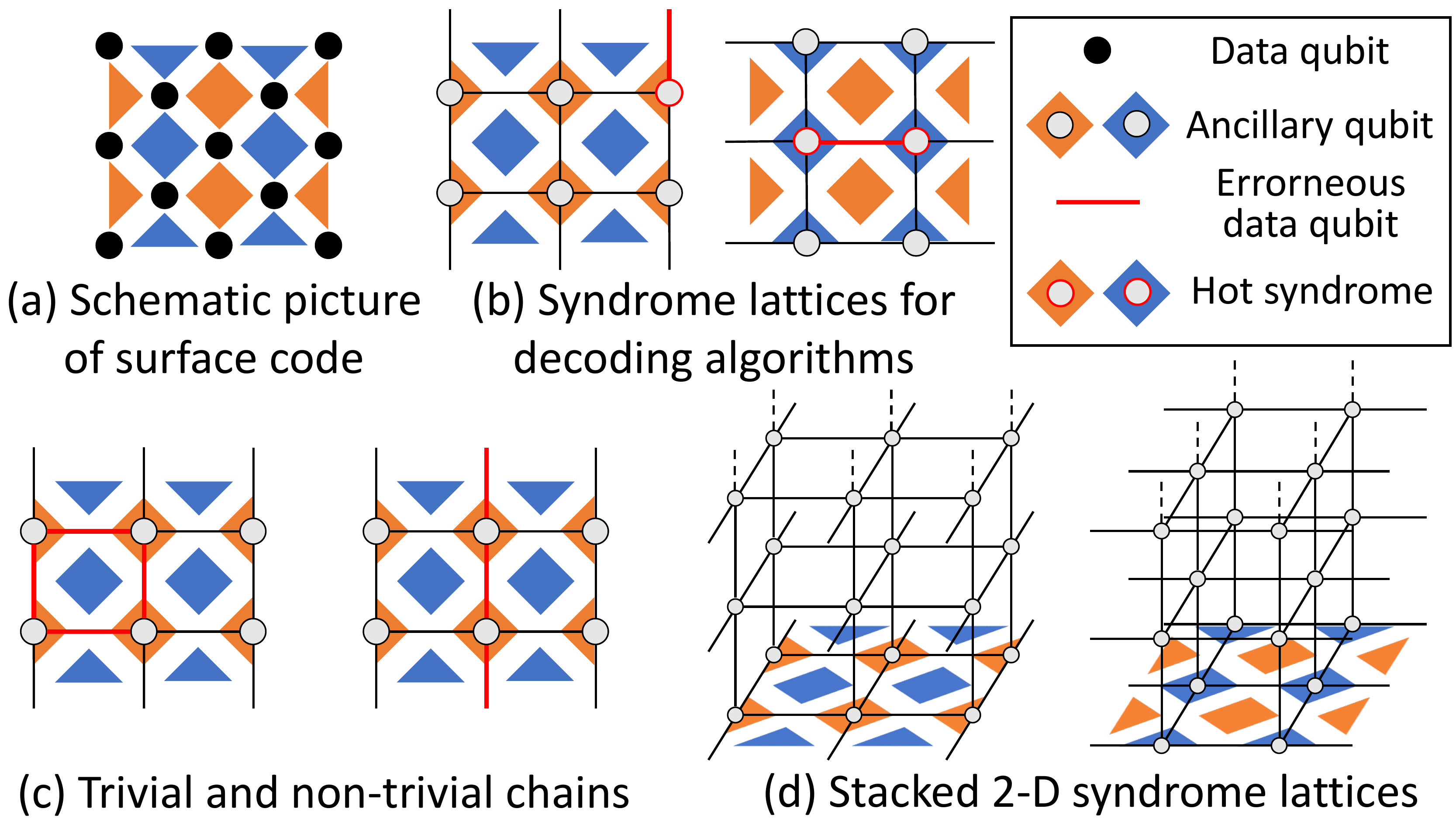}
    \vspace{-5mm}
    \caption{(a) Schematic picture of SC ($d = 3$). (b) Syndrome lattices generated from surface codes. (c) Examples of trivial and non-trivial chains of errors. (d) Stacked 2-D syndrome lattices.}
    \label{fig:surface_code_d3}
    \vspace{-3mm}
\end{figure}
To deal with both bit-flip (Pauli-$X$) and phase-flip (Pauli-$Z$) errors, QEC needs two types of stabilizer measurement: $X$ and $Z$. $X$- and $Z$-stabilizer measurements can detect the Pauli-$Z$ and -$X$ errors on the neighboring data qubits, respectively. Note that Pauli-$Y$ errors are considered to be a combination of bit- and phase-flip errors because the Pauli operators $X$, $Y$, and $Z$ have the following relationship: $Y = iXZ$.
Fig.~\ref{fig:surface_code_d3}\,(a) shows a schematic picture of the SC with code distance $d = 3$. Here, data and ancillary qubits for $X$-($Z$-) stabilizer measurements 
are represented as circles and blue (red) squares, respectively. $X$-($Z$-)stabilizer measurements detect Pauli-$Z$(-$X$) errors when the corresponding parity of the neighboring qubits is odd.

The decoding process of SCs is to estimate errors on physical qubits from outcomes of stabilizer measurements and commonly reduced to the following MWPM problem~\cite{fowler2012surface}.
Suppose a 2-D grid graph where its nodes and edges correspond to the syndrome values and data qubits, respectively. An example is shown in Fig.~\ref{fig:surface_code_d3}\,(b), where erroneous data qubits are represented as red edges, and detected syndromes are represented as nodes with red rims. The two lattices for $X$- and $Z$-stabilizer measurements have two distinct boundaries: smooth and rough boundaries, respectively. We construct a weighted complete graph where its nodes correspond to the detected syndromes or boundaries, and the weights of its edges are determined from the noise model. We find an MWPM of the complete graph, and the estimated errors are represented as edges between each pair of syndromes of the matching.
Error estimation success is determined by the properties of the occurred and estimated errors. If we estimate errors as a perfect matching, the resultant error chain consists of topologically trivial and non-trivial chains, as shown in Fig.~\ref{fig:surface_code_d3}\,(c). The odd number of non-trivial chains indicates the failure of error estimation.

Even when ancillary qubits also suffer from noise, we can reliably estimate errors by extending the decoding task to a 3-D lattice, {\it i.e.}, by considering stacked 2-D snapshots as shown in Fig.~\ref{fig:surface_code_d3}\,(d). A procedure to generate a 2-D snapshot is called a {\it code cycle}. A practical decoding algorithm must be capable of decoding these sequentially captured 2-D snapshots every code cycle.

%% file: NEOQECchap/bg_lattice_surgery.tex
\subsection{Lattice surgery}  
\label{subsec:bg_lattice_surgery}
During the computation, we need to perform a universal set of logical gates (Hadamard, CNOT, and $T$-gates) on logical qubits.
While logical Hadamard gates are straightforwardly implementable, the implementation of logical CNOT and $T$-gates only with neighboring physical operations is not trivial. 
The lattice surgery (LS) technique enables us to efficiently implement these two logical gates\cite{horsman2012surface}.

\begin{figure}[tb]
    \centering
    \includegraphics[width=\linewidth]{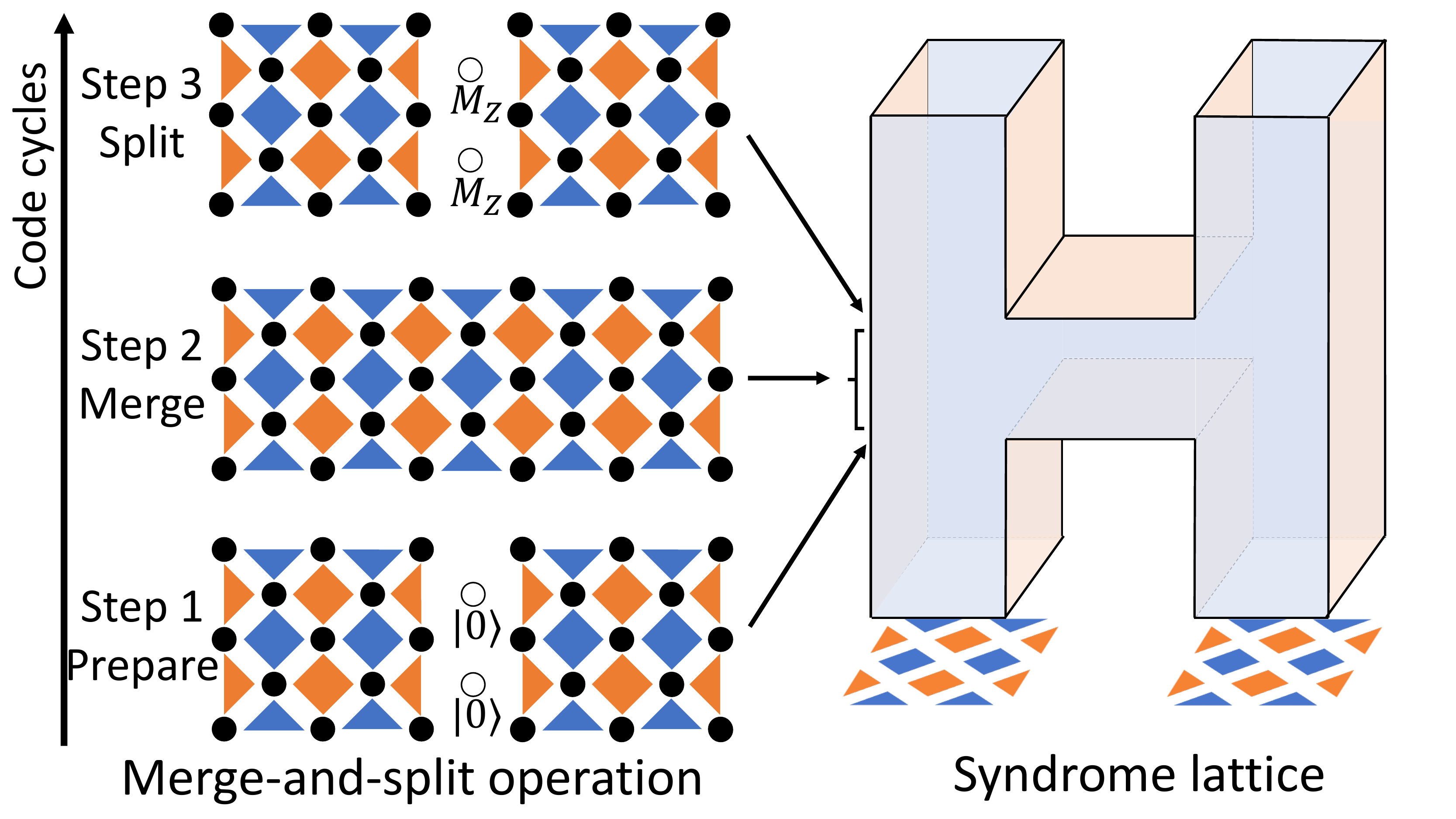}
    \caption{Process of logical Pauli-$XX$ measurements with LS (left side) and its syndrome lattice (right side).}
    \label{fig:lattice_surgery_example}
    %\vspace{-3mm}
\end{figure}

LS provides a way to implement logical Pauli measurements on multiple logical qubits via the following merge-and-split operations. The minimum example of the LS, a logical Pauli-$XX$ measurement on two logical qubits, is shown in Fig.~\ref{fig:lattice_surgery_example}. 
In this figure, (Step 1) we initialize all the sandwiched physical qubits to physical $\ket{0}$ states, (Step 2) two surface codes are merged by performing another set of stabilizer measurements and repeating them for $d$ code cycles, and (Step 3) split the merged codes into two planes by performing the original stabilizer measurements and measure all the sandwiched physical qubits on Pauli-$Z$ basis.
These operations are known to be equivalent to a logical Pauli-$XX$ measurement on the two logical qubits. The outcome of the logical Pauli measurement is calculated from the parity of the outcomes of Pauli-$X$ stabilizer measurements in the first cycle of LS. Since this procedure merges two rough boundaries, this operation is called a rough merge. The Pauli-$ZZ$ measurement can also be achieved in a similar way by merging smooth boundaries.
Note that a logical CNOT operation can be achieved through logical Pauli-$XX$ and Pauli-$ZZ$ measurements and feedback of logical Pauli operations; hence, the logical CNOT operation is implemented with two merge-and-split operations. 
In addition, a logical $T$-gate can be efficiently implemented with LS and gate teleportation with magic states. See Refs.~\cite{fowler2012towards,fowler2018low} for details of this formalism.

During the merge-and-split operations for rough boundaries, we obtain a stacked 2-D lattice to be decoded, as shown in the left side of Fig.~\ref{fig:lattice_surgery_example}. As in the case of the figure, a Pauli error connecting a boundary to another distinct one with the same color is undetectable with stabilizer measurements and modifies the logical states or flips the result of the logical measurement. We keep $d$ code cycles during the merge phase since the distance between two U-shaped red boundaries corresponds to the code cycles during the merge.

%% file: NEOQECchap/bg_SFQ.tex
\subsection{Single Flux Quantum logic \label{subsec:SFQ_intro}}

A single flux quantum (SFQ) logic\cite{RSFQ_review} is a pulse-driven digital circuit composed of superconductor devices.
It is one of the most practical technologies with the potential for utilization in next-generation computers because of its ultra-fast and low-power performance compared with CMOS. SFQ technology enables a low-level voltage pulse-driven switching that enables both fast switching ($\sim10^{-12}$s) and low-energy consumption ($\sim10^{-19}$ J per switching). Hence, it is feasible to improve the device's clock frequency to a few tens to one hundred GHz with this technology\cite{nagaoka2019a48ghz,2019nagaoka_isec}.
In this paper, we focus on rapid-SFQ (RSFQ) and its energy-efficient family, named energy-efficient RSFQ (ERSFQ)\cite{kirichenko2011zero} in particular, to design an ultra-low-power and high-speed QEC decoder in a cryogenic environment. 

\begin{figure}[tb]
    \centering
    \includegraphics[width=\linewidth]{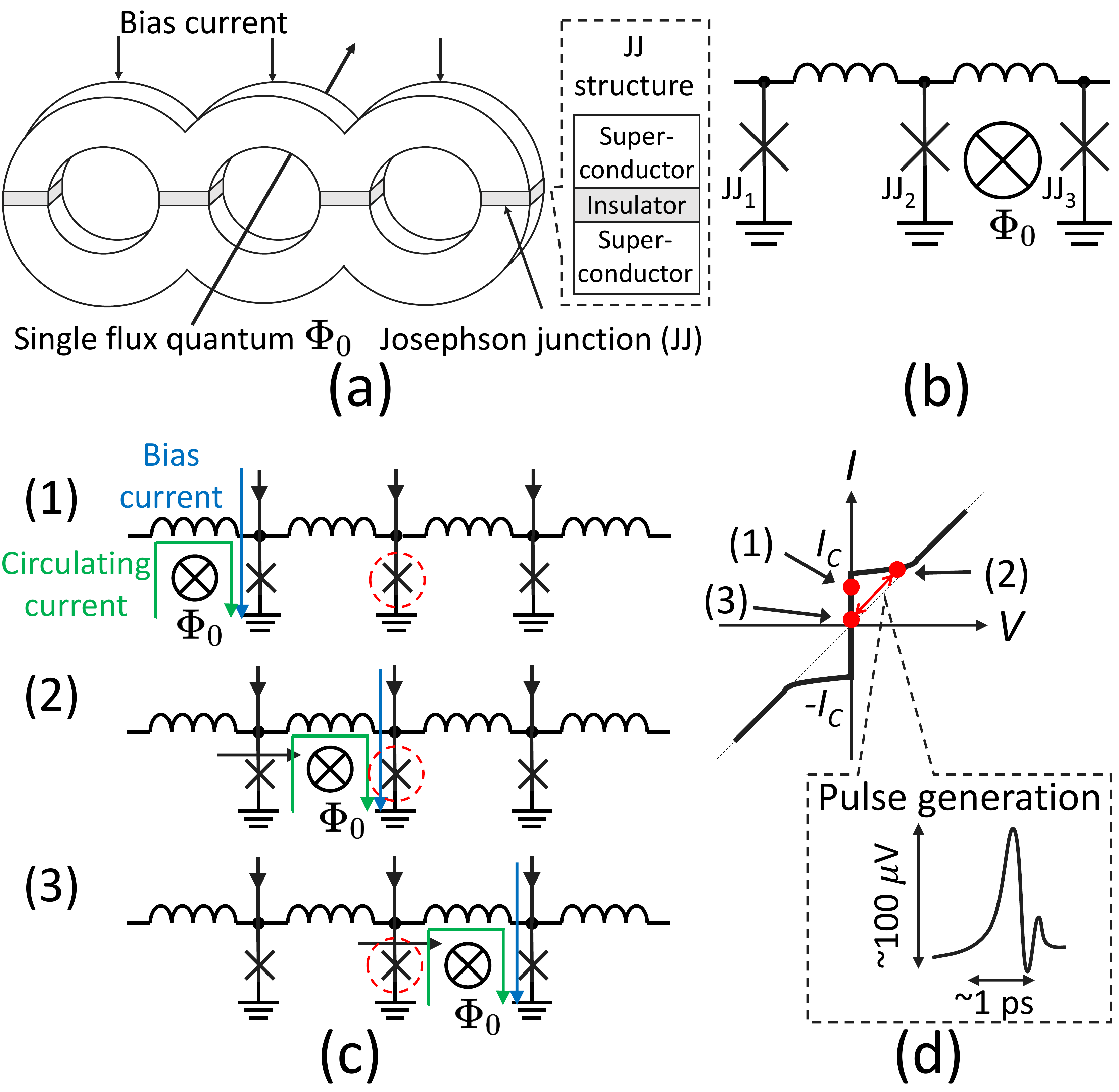}
    \caption{(a): Superconductor rings with single flux quantum and the structure of JJ. (b): Equivalent circuit of (a). (c): Transmission of single flux quantum on JTL. (d): Electrical characteristic of the JJ in the red dashed circle of (c).}
    \label{fig:SFQ_logic}
\end{figure}
Figure~\ref{fig:SFQ_logic}\,(a) shows the basic element of SFQ circuits. 
Information processing in SFQ circuits is performed with magnetic flux quanta stored in superconductor rings; the presence or absence of an SFQ in the ring represents a logical `1' or `0', respectively.
In SFQ circuits, an SFQ is stored or transferred using Josephson junctions (JJs) in the ring; the JJ has a ``superconductor--insulator--superconductor'' structure, as shown on the right side of Fig.~\ref{fig:SFQ_logic}\,(a). JJs in the ring act as a switching device similar to transistors in ordinary CMOS circuits.
Figure~\ref{fig:SFQ_logic}\,(b) shows the equivalent circuit diagram of the superconductor rings with JJs shown in Fig.~\ref{fig:SFQ_logic}\,(a).
The inductors and cross marks represent superconductor parts of the ring and JJs, respectively. The circle with a cross means an SFQ stored in the superconductor ring. 
Multiple rings connected in series form a Josephson transmission line (JTL), a type of SFQ wiring.

Figure~\ref{fig:SFQ_logic}\,(c) shows the transmission of an SFQ on a JTL. In step~(1) of Fig.~\ref{fig:SFQ_logic}\,(c), an SFQ is stored in the left ring and generates a circulating current in the ring. The sum of the circulating current of the SFQ (green arrow) and bias current (blue arrow) exceeds the critical current $I_{C}$ of the left JJ. In step~(2), the left JJ is switched, and the SFQ moves to the next ring. Then, the central JJ is similarly switched, and the SFQ moves to the right ring in step~(3).

Figure~\ref{fig:SFQ_logic}\,(d) shows the electrical characteristic of the JJ in the red circle of Fig.~\ref{fig:SFQ_logic}\,(c). 
In step~(1) of Fig.~\ref{fig:SFQ_logic}\,(c), only the bias current flows in the JJ, and it is smaller than the critical current $I_{C}$. In steps~(2) and (3), the sum of the circulating current and bias becomes greater than $I_{C}$. Then, the JJ is switched. While switching, the voltage of the JJ becomes zero, and it generates an impulse-shaped voltage pulse named an \textit{SFQ pulse}, as shown on the right side of Fig.\ref{fig:SFQ_logic}\,(d). 

An SFQ pulse has a quantized area $\Phi_0$, which is established by the following formula:
\begin{eqnarray} 
\int V(t)dt = \Phi_0 \approx 2.07 \times 10^{-15}~\text{Wb}.
\end{eqnarray}
The typical width and height of the SFQ pulse are a few picoseconds and a few hundred microvolts, respectively, as shown in Fig.~\ref{fig:SFQ_logic}\,(d).
The signal transmission based on SFQ pulses releases the SFQ circuits from the charging/discharging process required for the signal transmission of CMOS circuits, and it enables SFQ circuits to perform ultra-fast and low-energy information processing.

In the RSFQ circuit, the power consumption is calculated as follows:
\begin{eqnarray} \nonumber
\lefteqn{\text{(RSFQ power consumption)}_{\text{[W]}}} \\
&=& \text{(Bias voltage)}_{\text{[V]}} \times \text{(Bias current)}_{\text{[A]}}. \label{eq:RSFQ}
\end{eqnarray}
Most of the power of the RSFQ circuit is consumed statically, almost independent of the switching activities of JJs.

ERSFQ is a promising technology that completely excludes the static power dissipation of RSFQ\cite{kirichenko2011zero}. In ERSFQ, bias resistors are replaced with JJs, by which we can eliminate the large static power consumption in exchange for doubled dynamic power consumption by JJs. Because the JJ's dynamic energy consumption is roughly represented by bias-current $\times \Phi_0$ per switch, we can estimate the power consumption of ERSFQ on the basis of the bias current of RSFQ logic design and the power model of ERSFQ \cite{mukhanov2011energy} as follows:
\begin{align}
\lefteqn{\text{(ERSFQ power consumption)}_{\text{[W]}}} \\\nonumber
&=& (\text{Bias current})_{\text{[A]}} \times (\text{Frequency})_{\text{[Hz]}} \times \Phi_{0\text{[Wb]}} \times 2. \label{eq:ERSFQ}
\end{align}

%% file: NEOQECchap/02related_work.tex
\section{Related works\label{sec:related_work}}
\label{sec:BNN_related_work}

\subsection{Online decoder with SFQ circuits}
%There have been proposed several practical SC decoders. An architecture based on a UF-based decoding algorithm~\cite{das2020scalable} was recently proposed; they attracted much attention because of their accuracy and simplicity. Das {\it et al.}~\cite{das2020scalable} stated that fully pipelined hardware helps speed up the decoding process. 
Cryogenic computing, such as SFQ and Cryo-CMOS, has been actively studied to design peripherals of QC controllers~\cite{holmes2020nisq,2017micro_qureshi,qureshi_2017ACM}.
Holmes {\it et al.}~\cite{holmes2020nisq} proposed an algorithm named approximate quantum error correction (AQEC) where they devised a power-efficient and high-speed SFQ-based decoder for SC. Their implementation consists of multiple units corresponding to each data and ancillary qubit of an SC logical qubit to detect and correct errors by propagating simple signals between the units with a distributed processing scheme. Their implementation is capable of correcting Pauli errors on the data qubits. In 2021, we proposed the QECOOL decoder, an extension of AQEC to deal with measurement errors of ancillary qubits~\cite{ueno2021qecool}. We implemented an SFQ-based decoder that achieves a lower power consumption than AQEC. Furthermore, we proposed the QULATIS\cite{ueno2022qulatis} decoder by extending the QECOOL to deal with LS\cite{horsman2012surface}.

To cope with the measurement error, we also proposed the concept of online-QEC, where the stabilizer measurements and decoding processes are performed simultaneously, in contrast to the conventional batch-QEC, where the decoding process is done after all the measuring processes. The vast amount of error information in batch-QEC makes QEC latency worse, which causes errors to accumulate. Moreover, batch-QEC requires a large amount of memory to store error information proportional to the cube of code distance, which leads to decoders with a large hardware cost.
By contrast, online-QEC uses part of a syndrome lattice to decode SCs.
Therefore, it has the potential to be fast enough to avoid error accumulation and requires only a constant amount of memory independent of code distance.
Our previous SFQ-based decoders\cite{ueno2021qecool,ueno2022qulatis} used the online-QEC technique; however, due to their greedy nature and inability to process the entire 3-D syndrome lattice, they have lower accuracy than that of software decoders such as MWPM and the hardware efficient decoder based on the UF algorithm\cite{das2022afs}.
Therefore, we extend the previous SFQ-based decoders with NNs to build a lightweight and high-performance online decoder with SFQ circuits in this paper.

\subsection{Neural network-based decoders}

Decoders for SCs and other topological codes using NNs are widely studied and expected to achieve a near-optimal decoding performance by utilizing the correlation between Pauli $X$- and $Z$-errors\cite{torlai2017neural,maskara2019advantages,varsamopoulos2019comparing,sheth2020neural,meinerz2021scalable,bhoumik2021efficient,gicev2021scalable,overwater2022neural}.
When NN-based decoders were first proposed, many approaches trained NNs to predict the most probable physical errors or the required recovery operation for input syndrome values\cite{torlai2017neural}, and we refer to these approaches as \textit{end-to-end NN decoders}. However, generally, the result of a recovery operation inferred by end-to-end NN decoders is not necessarily a codeword. In addition, it has been reported that these end-to-end approaches are not scalable because the problems become more complex as the code distance increases, rendering training NNs more difficult\cite{meinerz2021scalable,gicev2021scalable}.

In contrast to end-to-end NN decoders, several studies use NNs combined with other decoding algorithms to build a hierarchical decoder\cite{meinerz2021scalable,gicev2021scalable,chamberland2022techniques}. These approaches use NNs at the first stage to correct small distance errors by a part of a syndrome lattice, and remaining long-distance errors are corrected with a conventional decoding algorithm, such as MWPM or UF, at the second stage. The second-stage decoder guarantees that its output is a codeword. 
In addition, the size of the syndrome input to NNs is small, independent of code distance, which keeps NNs size constant and their training easy. We call these approaches \textit{two-stage NN decoders}.

Meinerz \textit{et al.}~\cite{meinerz2021scalable} proposed a two-stage NN decoder that combined a fully-connected NN decoder with a UF decoder that focuses on decoding problems of 2-D and 3-D Toric Code~\cite{KITAEV20032}. The decoding process of Toric code is similar to that of SC except for the boundary condition of logical qubits. Their work achieved scalability and high accuracy by dividing the decoding process into two stages: NN preprocessing for local error corrections and longer-range error corrections with the UF decoder. For the 2-D (3-D) Toric code, the $X$- and $Z$-syndromes within the square (octahedral) $L$ on a side surrounding a certain data qubit are input to the NN decoder in the first stage, where $L$ is a constant independent of code distance $d$. The NN decoder outputs the most probable error among $\{I, X, Y, Z\}$ on the central data qubit of the square (octahedral). This procedure is performed for all data qubits, and the error chains smaller than $L$ are expected to be corrected. The remaining errors are corrected in the second stage with the UF decoder.

\begin{figure}[tb]
    \centering
    \includegraphics[width=\linewidth]{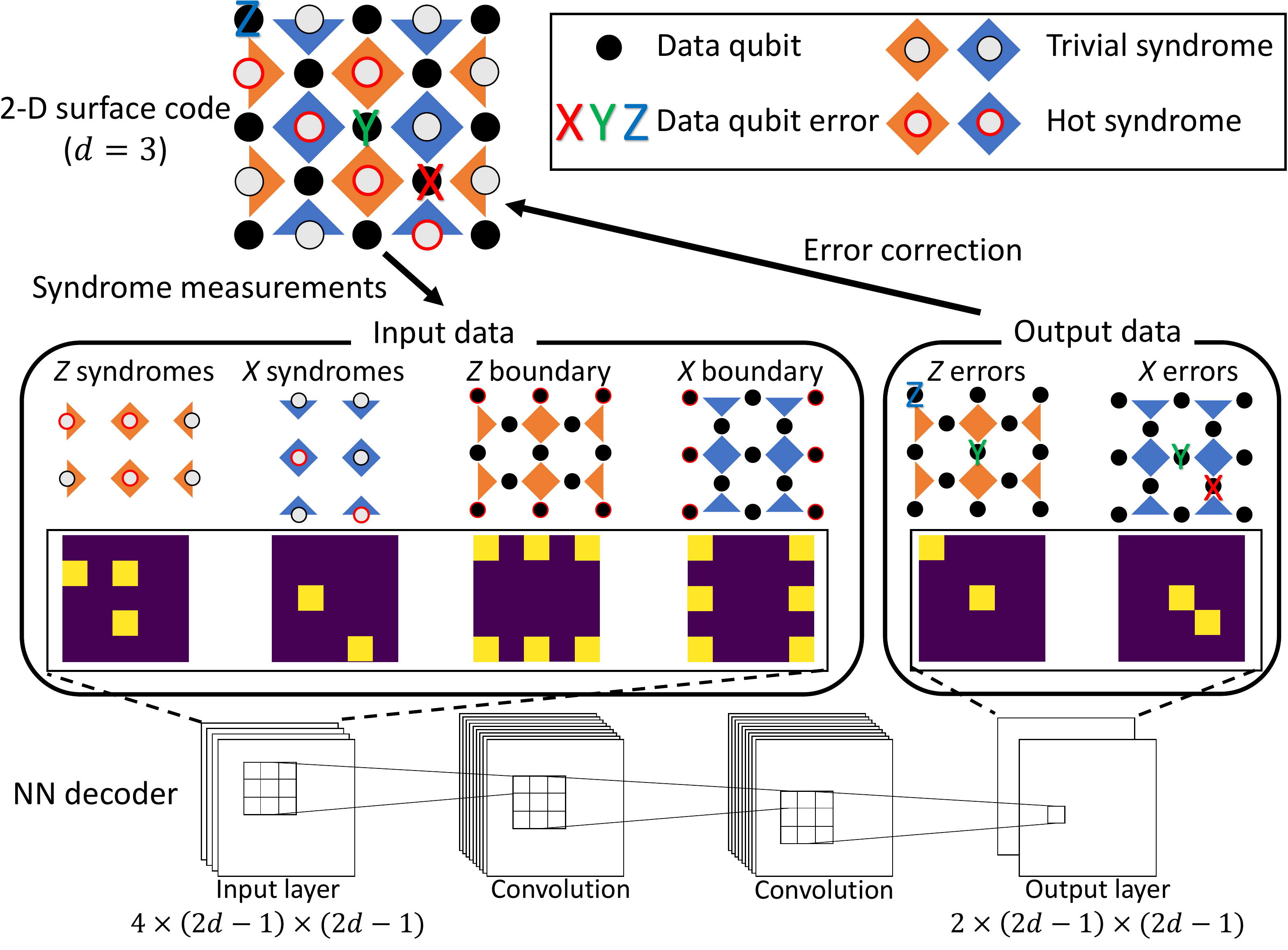}
    \caption{Overview of the NN decoder of the previous work\cite{gicev2021scalable}.}
    \label{fig:NN_previous_2D}
\end{figure}

Gicev \textit{et al.}~\cite{gicev2021scalable} also proposed a two-stage NN decoder that combined an NN with a hard-decision renormalization group (HDRG) decoder\cite{watson2015fast}.
They focused on decoding SCs under depolarizing noise on data qubits without measurement errors of ancilla qubits.
Although their approach is similar to the previous one by Meinerz's group\cite{meinerz2021scalable}, the NN used in the first stage is different; their NNs consist of only convolutional layers, whereas those of Ref.~\cite{meinerz2021scalable} have only fully-connected layers.
Figure~\ref{fig:NN_previous_2D} shows the overview of their NN-based decoder for 2-D SCs with an example of code distance $d = 3$. 
As shown in the figure, their NN decoder consists of several convolutional layers.
The fully convolutional decoder uses a four-channel input with binary values; the first two channels correspond to the $X$- and $Z$ syndromes, and the last two channels correspond to the $X$- and $Z$- boundary information. 
The input value of 1 represents a hot syndrome or a boundary data qubit, as shown in the example of Fig.~\ref{fig:NN_previous_2D}.
They explicitly use boundary information of logical qubits as input of NNs, which leads to the robustness of the NNs to boundary changes, including LS\cite{horsman2012surface} and braiding\cite{fowler2012surface}. 
The decoder has several hidden convolutional layers between the input and output layers. The output shape of each layer is set to the same shape as the input except for channel numbers. 
The output layer is a convolutional layer with two kernels, and its activation function is a sigmoid function, whose return value is between $0.0$ and $1.0$.
Therefore, the output of the decoder consists of two channels, and each channel represents the probabilities of $X$ or $Z$ errors on the associated data qubits, respectively. 
Finally, all the data qubits with output values greater than $0.5$ are corrected.

Although their decoder achieved high accuracy with 2-D SC lattices, it is not clear whether the decoder would work well in a situation with measurement errors of ancilla qubits. Moreover, no hardware implementation of the decoder has been presented, and whether the decoder satisfies all of the other three requirements of practical decoders except for accuracy, has yet to be evaluated.

In this paper, we extend this approach to be suitable for online decoding of 3-D complex lattices and combine it with the QECOOL and QULATIS decoder to achieve a new online decoding algorithm with higher decoding accuracy. In addition, we design an ultra-low-power and fast NN decoder with SFQ circuits and show that it meets all of the four requirements.

%%%%%%%%%%%%%%%
Independently to this work, Chamberland \textit{et al.}~\cite{chamberland2022techniques} proposed a similar idea to this work, combining an NN decoder with a conventional decoder. They used fully convolutional networks to design a two-stage NN decoder that worked well for circuit-level noise on SCs of arbitrary size and estimated its resource cost on the basis of FPGAs. The main difference to our work is that we use quite simple CNNs and a binarization technique to achieve a low-latency and low-power NN decoder in a cryogenic environment.

%% file: NEOQECchap/03proposed.tex
\section{NN-based two-stage decoder\label{sec:BNN_proposed}}
In this section, we propose the NEO-QEC algorithm by extending the QECOOL and QULATIS decoders with the existing NN-based two-stage decoder\cite{gicev2021scalable}.
We construct the first stage NN-based decoder with the decoding capability for 3-D SC lattices by extending the previous work\cite{gicev2021scalable} and use the QECOOL or QULATIS as the non-NN decoder in the second stage.

\subsection{Construction of NN-based decoder for 3-D SCs\label{subsec:decoding_algorithm}}

\begin{figure}[tb]
    \centering
    \includegraphics[width=\linewidth]{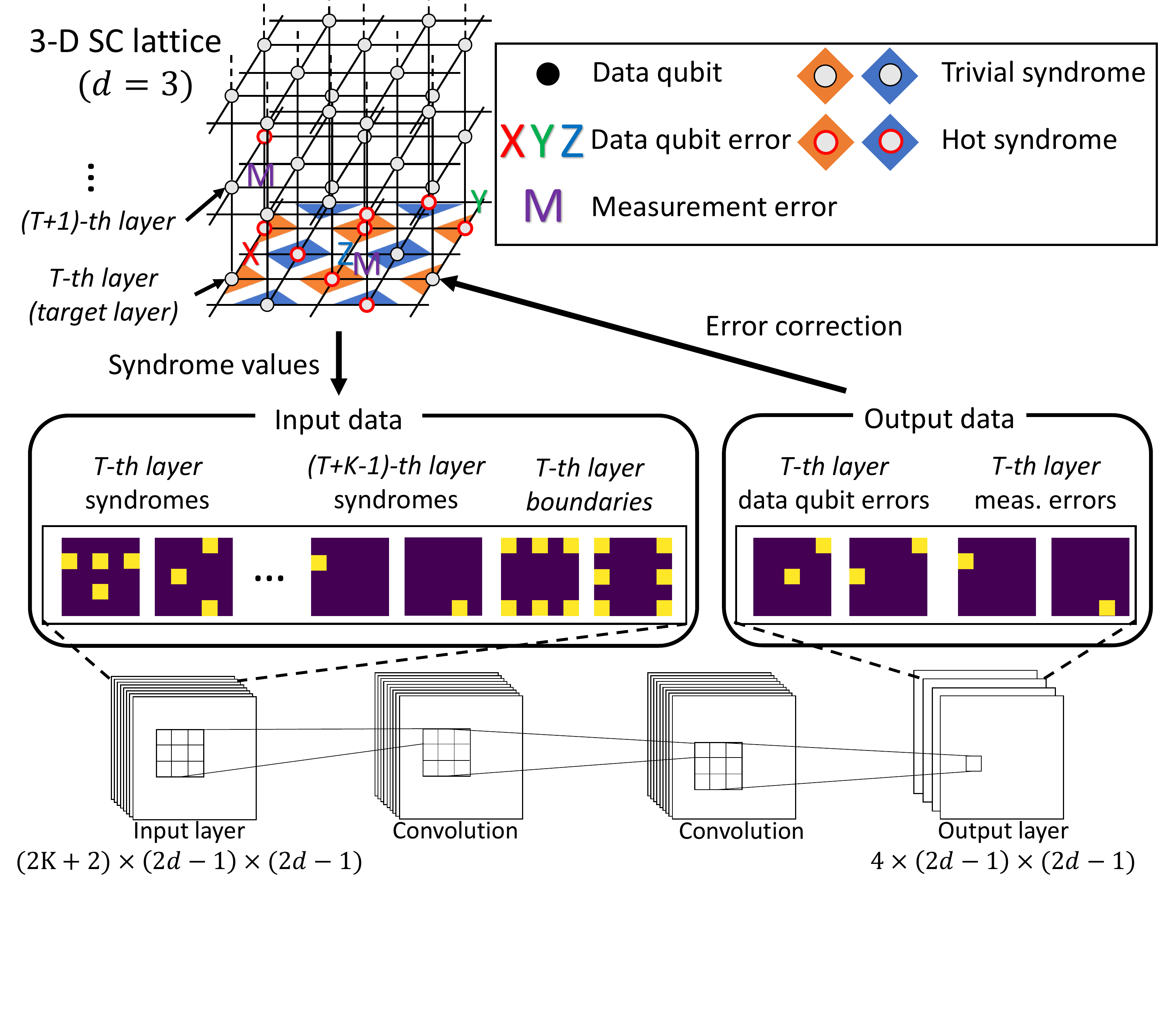}
    \vspace{-15mm}
    \caption{Overview of our NN-based decoder.}
    \label{fig:NN_3D}
\end{figure}

We extend the NN decoder of the previous work\cite{gicev2021scalable} to handle measurement errors.
Figure~\ref{fig:NN_3D} shows an overview of our NN-based decoder for 3-D SCs with an example of code distance $d = 3$. 

Decoding SCs with measurement errors requires measuring ancilla qubits multiple times and stacking each result temporally to create a 3-D SC lattice. 
To simplify the process of the NN decoder, we train the NN to infer errors on a specific layer (called the \textit{target layer}) on the 3-D syndrome lattice rather than the entire 3-D lattice.
We introduce an integer parameter $K$, which represents the number of layers of the syndrome lattice used as an input for the NN decoder. 
In other words, the input of the NN decoder is a $K$-layer syndrome lattice that includes the target layer as the lowest layer. In addition, the input includes the boundary information of the target layer, similar to the previous study~\cite{gicev2021scalable}, resulting in a $2K+2$ channel input.
The decoder is a fully convolutional network with several hidden layers, and each convolutional layer has several constant-size kernels independent of code distance $d$. Its output has four channels, and each channel corresponds to $X$- and $Z$-errors on data qubits and measurement errors on $X$- and $Z$-ancilla qubits, respectively. Table~\ref{tab:differences} summarizes the differences between our method and that proposed in the previous study.
Note that our NN decoder is capable of processing 3-D SC lattices with various code distances and shapes.

To decode a 3-D SC lattice of height $d$, the NN decoder needs to perform inference $d$ times in total, changing the target layer from the bottom to the top. The NN decoder requires only $K-1$ additional measurement processes to infer errors on a target layer and does not have to wait until all measurement processes are performed, which is suitable for online-QEC as explained in Section~\ref{subsec:online_decoding_with_NN}.

\begin{table}[tb]
\caption{Comparison of NEO-QEC and the method proposed in the previous work\cite{gicev2021scalable}.}
\label{tab:differences}
\centering
\scalebox{0.8}{
\begin{tabular}{|c|c|c|c|c|}\hline
& Lattice & \begin{tabular}[c]{@{}c@{}}Meas. \\ error\end{tabular} & Input channels & Output channels \\ \hline
\begin{tabular}[c]{@{}c@{}}Previous work \\\cite{gicev2021scalable}\end{tabular} & 2-D SC & No & 4 & 2 \\ \hline
\begin{tabular}[c]{@{}l@{}}NEO-QEC\\ (this work)\end{tabular} & 3-D SC & Yes & $2K+2$  & 4 \\ \hline
\end{tabular}
}
\end{table}

\subsection{Binarization of NN-based decoder\label{subsec:binarization}}

In general, NNs require heavy computational resources. Therefore, many techniques, such as knowledge distillation\cite{gou2021knowledge} or pruning\cite{liang2021pruning}, are proposed to develop fast and lightweight NNs with slight accuracy degradation. Quantization is a technique that converts 64- or 32-bit floating numbers in the model parameters of NNs to 8-bit integers or lower bitwidth representations\cite{Gholami2021survey}. This technique enables for compact NN models and has shown tremendous and consistent success in implementing high-speed and low-power NNs\cite{jouppi2017datacenter}.

Binarization of NNs is the most extreme quantization method, where 1-bit values represent both weights and input data of NNs\cite{hubara2016binarized,Rastegari2016XNORNetIC,liu2018bi,zhu2020XOR}, thereby drastically reducing the memory requirement compared with floating-point representations.
In addition to the memory space advantages, binary (1-bit) operations can often be computed efficiently with bit-wise arithmetic and achieve significant acceleration.

In XNOR-net\cite{Rastegari2016XNORNetIC}, one of the implementations of binarized NNs, the costly floating-point multiplications on NNs are replaced with lightweight XNOR operations followed by bit counters. XNOR-net requires binarization of input data and model parameters, which generally leads to significant accuracy degradation. However, since each element of input and output in the proposed NN decoder is originally binary data, the effect of binarization on accuracy is expected to be less significant.
In this work, we implement an NN decoder with XNOR-net-based binarized convolutional NN.

\subsection{Online decoding with NN decoder\label{subsec:online_decoding_with_NN}}

\begin{figure}[tb]
    \centering
    \includegraphics[width=\linewidth]{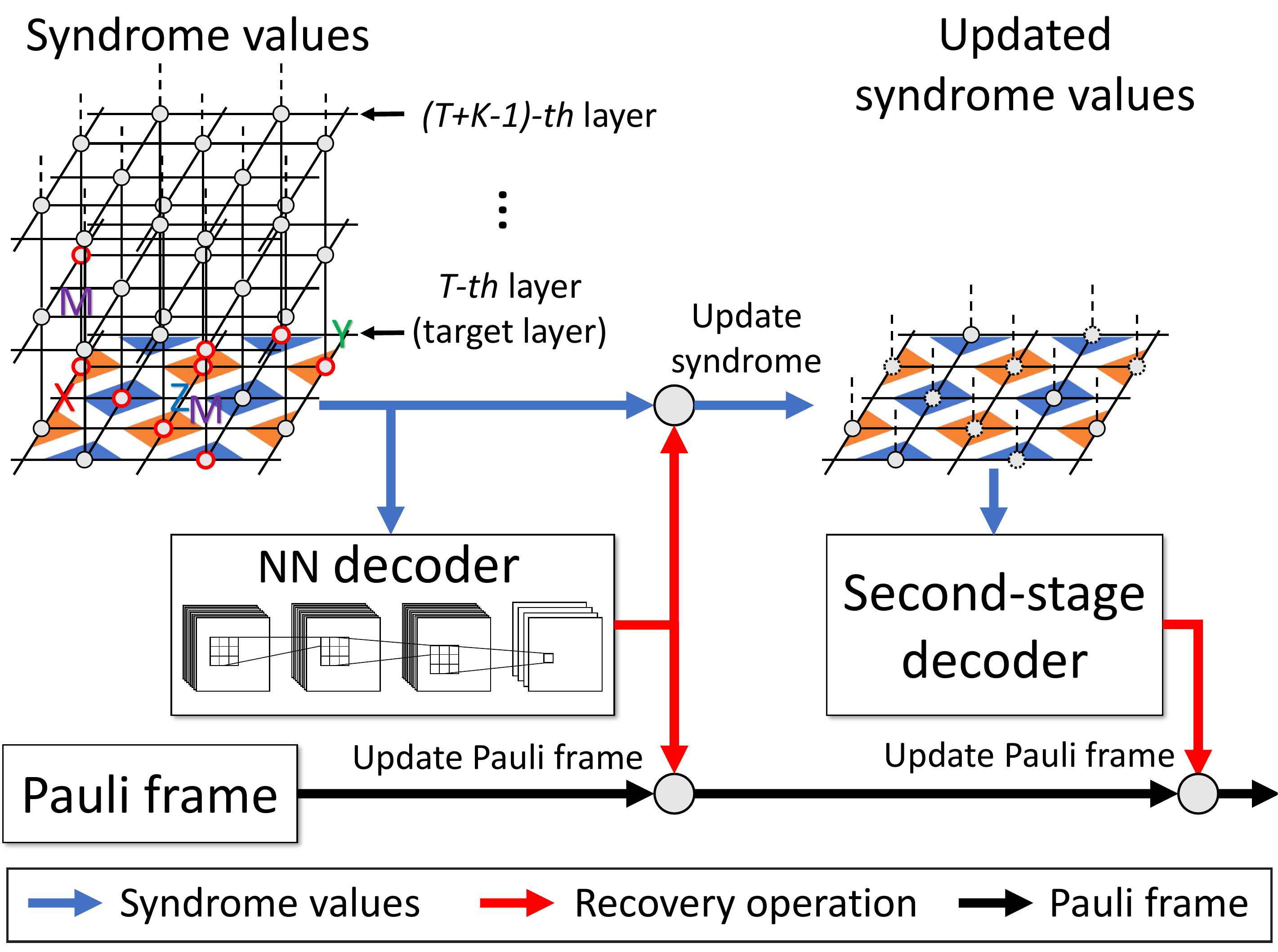}
    \caption{Workflow of the NEO-QEC decoder.}
    \label{fig:online_architecture_with_NN}
\end{figure}

We combine the NN decoder with the QECOOL or QULATIS decoder as the NEO-QEC decoder architecture for online SC decoding. Figure~\ref{fig:online_architecture_with_NN} shows the workflow of our NEO-QEC decoder.
Note that the second-stage decoder is not limited to QECOOL and QULATIS, and any online SC decoders are suitable.
%The system will output the estimated errors in accordance with the matching result, which are sent to the unit called Pauli frame~\cite{riesebos2017} 

We explain the online decoding workflow of the NEO-QEC algorithm as follows. The SC measurement process is repeated at a specific interval, and syndrome values are extracted every code cycle to build an SC lattice until the logical measurement.
Suppose that the bottom layer of the lattice is set to the target layer as shown in Fig.~\ref{fig:online_architecture_with_NN}. After the first $K$ times measurement, the syndrome values for $K$ layers of the syndrome lattice and the boundary information of the target layer are input to the NN decoder.
The NN decoder infers errors on the target layer, and we update the syndrome of the target layer and the Pauli frame\cite{riesebos2017pauli}\footnote{The basic idea of Pauli frames is to track error correcting operations represented by Pauli gates as classical information rather than applying them to physical qubits by leveraging the fact that error correcting operations are commutative with logical Pauli and Clifford operations.} on the basis of the inferred recovery operation as shown in the figure.
The updated syndrome values are saved in the buffer of the second-stage decoder.
After the subsequent measurement, the target layer is changed to the next layer from the previous one, and the same aforementioned procedure is performed.
Then, the buffer of the second-stage decoder stores the updated syndromes of the first and second layers.
After the aforementioned procedure is repeated several times, if the second-stage decoder stores sufficient syndrome layers to perform online decoding, \textit{e.g.}, the number of syndrome layers exceeds the threshold value denoted as $th_v$ in QECOOL or QULATIS, the second-stage decoder decodes the stored syndrome values; otherwise, the second-stage decoder waits for the completion of the NN decoder process. Thus, the second-stage decoder perform with a $K$-cycles delay for the NN decoder.

To achieve online processing of the whole architecture of NEO-QEC, the processing times of both NN and the second-stage decoders for a single layer of the SC lattice must be less than the code cycle (\textit{e.g.}, approximately $1~\mu$s for a QC with superconducting qubits).

%The $DI_W$ is calculated for the extracted syndrome values in our architecture.
%If the value of $DI_W$ is greater than a predetermined threshold, the syndromes are stored in the buffer without any preprocessing.
%Once the $DI_W$ exceeds the threshold, all subsequent syndrome values are input to the NN decoder in every code cycle until the logical measurement.
%Based on the output of the NN decoder, data qubits are corrected, and the syndrome values are updated and saved in the buffer.
%The second-stage decoder then decodes the syndrome values stored in the buffer after each processing of the NN decoder or the measurement process.
%Note that we assume the QECOOL decoder as the second-stage decoder. However, other online decoders are also suitable.

%% file: NEOQECchap/04implementation_counter.tex
\section{Hardware implementation\label{sec:BNN_hardware_section}}
In this section, we design a neural processing unit (NPU) with SFQ digital circuits. Although several SFQ arithmetic units supporting floating-point operations have been proposed\cite{hara2009design,kainuma2011design,XizhuPENG2014,peng2015fpu,yamanashi2019measurement}, they consume a large amount of power. Thus, NPUs with floating-point operations are not suitable for the NEO-QEC decoder.
As described in Section~\ref{subsec:binarization}, the binarization technique based on XNOR-net\cite{Rastegari2016XNORNetIC} is an effective approach in this study because of its simple structure consisting of SFQ-friendly bit-wise arithmetic. Therefore, we design an NPU based on XNOR-net.

\subsection{XNOR-net-based NN with SFQ logic}
\begin{figure}[tb]
    \centering
    \includegraphics[width=\linewidth]{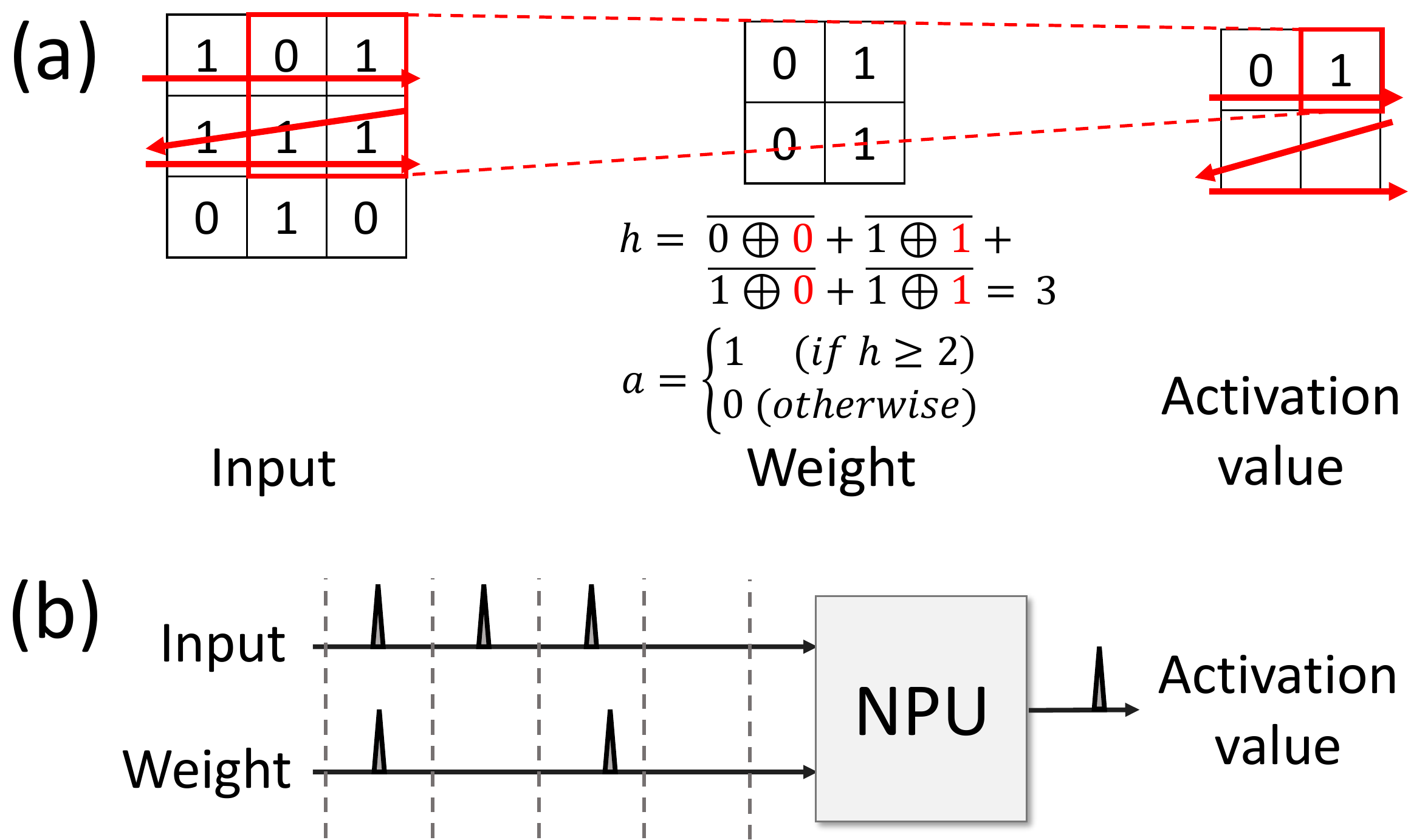}
    \vspace{-3mm}
    \caption{(a): Example of convolutional calculation with XNOR-net. (b): SFQ implementation of (a).}
    \label{fig:SFQ_XNOR_calc}
\end{figure}
Figure~\ref{fig:SFQ_XNOR_calc}\,(a) shows an example of an XNOR-based convolutional layer. Here, inputs, weights, and activation values are represented by 0 or 1, while the original work uses -1 or 1\cite{Rastegari2016XNORNetIC}. Multiply-and-add operations to calculate activation values are replaced with ``XNOR-and-add operations'', where the activation value is 1 if the majority of the XNOR results are 1, and 0 otherwise.

Figure~\ref{fig:SFQ_XNOR_calc}\,(b) illustrates an SFQ implementation of an XNOR-net-based convolutional calculation with the example of (a). As shown in the figure, binarized input and weight values are converted to bit-serial pulse trains where the absence and presence of a pulse at each time interval represent 0 and 1, respectively. The NPU receives pulse trains of input and weight values and outputs a pulse if the activation value is 1.

\subsection{Design of NPU with SFQ circuit}
\begin{figure}[tb]
    \centering
    \includegraphics[width=\linewidth]{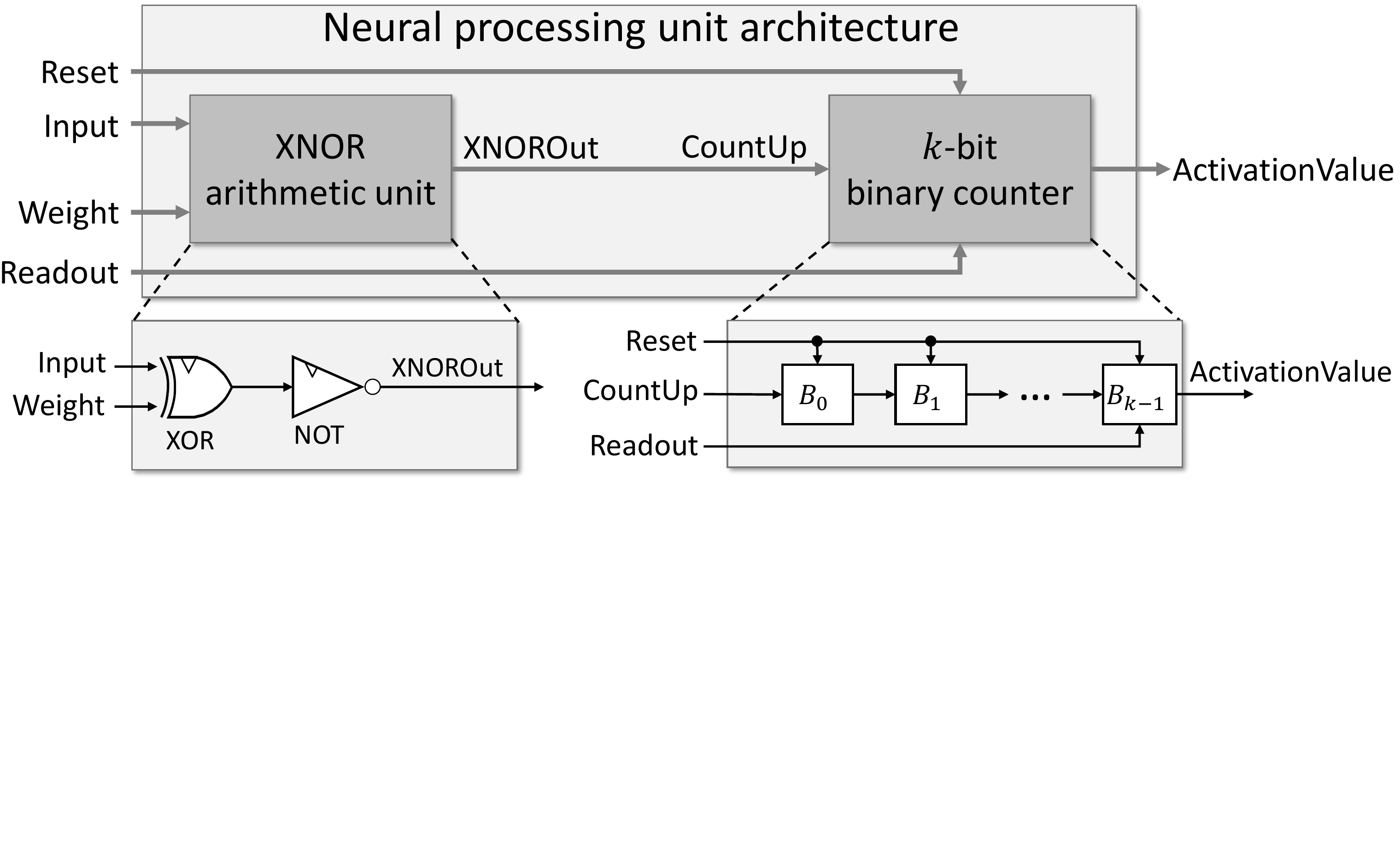}
    \vspace{-30mm}
    \caption{Overview of XNOR-net-based NPU for SFQ design.}
    \label{fig:XNOR_NPU_overview}
\end{figure}
As shown in Fig.~\ref{fig:XNOR_NPU_overview}, our SFQ-based NPU consists of an XNOR arithmetic unit and $k$-bit binary counter.
The detailed descriptions of the submodules are as follows.
\begin{enumerate}
    \item \textbf{XNOR arithmetic unit:} The XOR gate receives pulses representing input data and NN weight simultaneously. The negation of its output is used as the input signal of the $k$-bit binary counter. Note that the XNOR arithmetic unit outputs ``XNOROut'' with one cycle delay to the inputs because all the SFQ logic gates used in the XNOR arithmetic unit are clocked gates.
    \item \textbf{$k$-bit binary counter:} 
    The $k$-bit binary counter consists of $k$ T flip-flops (TFFs) and one D flip-flop (DFF) connected in series and counts the number of ``XNOROut'' pulses from the XNOR arithmetic unit to calculate the activation value. The DFF holds the output pulse of $B_{k-1}$ and sends it as ``ActivationValue'' after receiving the ``Readout'' pulse. Note that the initial value of the counter must be adjusted so that $B_{k-1}$ outputs a pulse after the counter receives half of the total number of input pulses.
\end{enumerate}
The bit length $k$ of the binary counter is determined by the maximum number of XNOR operations in a convolutional operation to be performed; the binary counter must be sufficiently large to count half the number of XNOR operations in a single convolution operation. In this paper, we target small- and medium-scale CNNs and evaluate decoder performance on the basis of the NPU with a 9-bit binary counter.

\subsection{NPU implementation based on RSFQ logic}
\begin{table}[t]
\centering
\caption{Summary of SFQ logic elements used in this paper.\label{tab:BNN_SFQ_logic_elements}}
\begin{tabular}{|l|c|c|c|c|} \hline
Cell                          & JJs &\begin{tabular}[c]{@{}c@{}}Bias current\\ (mA)\end{tabular} & \begin{tabular}[c]{@{}c@{}}Area\\ ($\mu \mathrm{m}^2$)\end{tabular}& \begin{tabular}[c]{@{}c@{}} Latency\\ (ps)\end{tabular}\\ \hline \hline
%Splitter                      & 3        & 0.300 & 900   & 4.3   \\
%Merger                        & 7        & 0.880 & 900   & 8.2   \\
%1:2 switch                    & 33       & 3.464 & 8100  & 10.5    \\
DFF                           & 6        & 0.720 & 900   & 5.1   \\
TFF                           & 13       & 0.808 & 3600   & 7.3   \\
%nondestructive readout (NDRO) & 11       & 1.112 & 1800  & 6.4    \\
%resettable DRO (RD)           & 11       & 0.900 & 1800  & 6.0 \\ 
%Triple-output DRO (D3)        & 17       & 1.241 & 3600  & 6.8   \\ 
XOR                           & 11       & 1.068 & 3600  & 6.5 \\ 
NOT                           & 11       & 0.848 & 3600  & 6.5 \\  \hline
\end{tabular}
\end{table}

We designed the NPU with a 9-bit binary counter based on an RSFQ cell library \cite{detail_of_cell_library_ADP2} developed for a niobium nine-layer, 1.0-$\mu$m fabrication technology \cite{Shuichi_NAGASAWA2014,Akira_FUJIMAKI2014}.
Table~\ref{tab:BNN_SFQ_logic_elements} summarizes the SFQ logic gates used in this paper. Because the essential element of SFQ that affects power consumption and hardware cost is the JJ, Table~\ref{tab:BNN_SFQ_logic_elements} shows the number of JJs for each gate and the bias current required for operation. The operating temperature and designed supply voltage are 4-K and 2.5~mV, respectively.

We used the Josephson simulator (JSIM)\cite{jsim}, a SPICE-level simulator, to verify the functionality of the designed NPU with a 9-bit binary counter and evaluate its latency.
Table~\ref{tab:BNN_JJcounts} shows the total number of JJs, total bias current, and latency of each submodule.
The NPU consists of 151~JJs in total, and its total bias current is $11.2$~mA. The maximum delay of the designed circuit is 13.8~ps, resulting in a maximum operating frequency of approximately 70~GHz. Note that Tab.~\ref{tab:BNN_JJcounts} shows the design results before detailed routing and does not include the wiring cost. Therefore, the power consumption and delay are estimated to be lower than the actual circuit to be designed.
\begin{table}[tb]
\centering
\caption{Total number of logic elements, number of JJs, and latency of each XNOR-net-based NPU\cite{Rastegari2016XNORNetIC} based on the AIST 10-kA/cm$^2$ ADP cell library\cite{detail_of_cell_library_ADP2}.}
\label{tab:BNN_JJcounts}
\begin{tabular}{|l|c|c||c|} \hline
Cell       & \begin{tabular}[c]{@{}c@{}}Binary \\counter \\ (9-bit)\end{tabular} & \begin{tabular}[c]{@{}c@{}}XNOR \\arithmetic \\ unit\end{tabular} & \begin{tabular}[c]{@{}c@{}}Total\\ (9-bit)\end{tabular}  \\ \hline
%Splitter              &      & 2    & 20        \\
%Merger                & 11   &      & 11         \\
DFF                   & 1    &      & 1          \\
TFF                   & 9    &      & 9         \\ 
%D3                    & 7    &      & 7         \\ 
XOR                   &      & 1    & 1         \\ 
NOT                   &      & 1    & 1          \\ \hline \hline
%Wire                 &      & 22    & 1132         \\ \hline \hline
Total JJs             & 123  & 28   & 151      \\
Total bias current (mA)& 7.99  & 3.23   & 11.2 \\ 
Latency (ps)          & 6.5    & 7.3    & 13.8   \\ \hline  
\end{tabular}
\end{table}

%% file: NEOQECchap/05evaluation.tex
\section{Evaluation}
\label{sec:BNN_evaluation}
\subsection{Evaluation setup}
In this section, we numerically evaluate the QEC performance of the NEO-QEC algorithm for single logical qubit protection and an LS procedure. 
We assume a depolarizing noise model, where Pauli-$X$, -$Y$, and -$Z$ errors occur on each data and ancillary qubit independently in each QEC cycle, which is called the phenomenological noise model. 
Since a Pauli-$Y$ error is considered as the combination of Pauli-$X$ and -$Z$ errors, there is a correlation between decoding Pauli-$X$ (bit-flip) and Pauli-$Z$ (phase-flip) errors.
%Its probability distribution is described as follows: 
%\begin{align*}
%    [p_I, p_X, p_Y, p_Z] = [1-p, p/3, p/3, p/3], 
%\end{align*}
%where $p$ is an error probability per physical qubit.
%Note that while actual noise would be locally correlated and coherent, the surface codes are expected to correct these errors with a modest performance degradation~\cite{bravyi2018correcting,fowler2014quantifying}. Thus, this evaluation captures the essential performance of our proposal.

We repetitively sample error patterns for a given physical error rate, simulate the propagation of errors and decoding procedure, and evaluate the probability of logical failures.
Although a complete simulation of quantum circuits requires time that scales exponentially with the number of qubits, we can efficiently simulate the propagation of Pauli errors because we assume Pauli errors and QEC circuits consist of Clifford gates and Pauli measurements~\cite{gottesman1998heisenberg}.

\subsection{Training\label{subsec:trainig_BNN}}

\begin{table}[tb]
\centering
\caption{Training parameters of NNs.\label{tab:train_parameter}}
\scalebox{0.9}{
\begin{tabular}{|l||l|} \hline
Parameter                             & Configuration       \\ \hline \hline
Epochs                                & 100                 \\
Training dataset size $N_{train}$     & Idling: $1 \times 10^6$, LS: $2 \times 10^5$     \\
Validation dataset size $N_{val}$     & $1 \times 10^5$     \\
Training error rate $p_{train}$       & Idling: 0.04, LS: 0.006                \\
Training lattice size $d_{train}$     & 9                   \\
Input syndrome layers $K$             & 3, 4, 5             \\
Input layer shape ($C, H, W$) & $ (2K+2, 2d_{train}-1, 2d_{train}-1)$                     \\
Output layer size ($C, H, W$) & $ (4, 2d_{train}-1, 2d_{train}-1)$\\
\# of total conv layers  & 2, 3 \\ 
\# of hidden layer channels       & 9, 16 \\
Kernel size                           & $5\times5$, $7\times7$ \\
Optimizer                             & ADAM                        \\
Learning rate                         & FP32: 0.01, BNN: 0.001 \\
Loss function                         & Binary cross-entropy              \\ \hline
\end{tabular}
}
\end{table}

We use standard supervised learning techniques to train NNs consisting of two or three convolutional layers for single logical qubit protection or an LS procedure, and the training parameters are shown in Tab.~\ref{tab:train_parameter}. 
As explained in Section~\ref{sec:BNN_proposed}, we use $K$ layers of a SC syndrome lattice and $X$- and $Z$-boundaries information of the target layer as input data, which consists of $2K+2$ channels. 
We use error information of each data qubit and measurement process on the target layer as label data. 
For the idling operation of a single logical qubit, the training data set is generated by simulating a depolarizing noise model with error probability $p_{train}$ on a distance-$d_{train}$ SC during $d_{train}$ code cycles.
For the merge-and-split operation, we generate the training data set by simulating the merge-and-split operation with two distance-$d_{train}$ logical qubits during $3d_{train}$ code cycles. 
In other words, the pair of input and label data is generated by performing the opposite operation of decoding, where the syndrome values are calculated from the randomly provided physical error information. 
This procedure is performed very efficiently, and we can create a large dataset to reduce the chance of overfitting. 
Note that NN structures are common for the two different tasks of single logical qubit protection and LS, while the training data sets are generated for each task.
The starting learning rate is shown in Tab.~\ref{tab:train_parameter}, and we decrease it exponentially during training, and its schedules are optimized by hand. 
The NNs are built with the PyTorch v1.8 platform\cite{NEURIPS2019_9015}.

\subsection{NN-based decoder\label{subsec:eval_NN_preprocessing}}

First, we show the error correction performance of the NEO-QEC algorithm for the idling operations of a single logical qubit during $d$ code cycles assuming the existence of measurement errors and compare it with the existing decoding algorithms. Here, the NNs of NEO-QEC are trained with 32-bit precision, and no quantization technique is used.

Table~\ref{tab:BNN_model_search} shows the accuracy of NEO-QEC with various configurations of the NN decoder.
Compared with the QECOOL, pseudo thresholds of NEO-QEC improve in all cases verified here.
Although the parameter size of the NN decoder varies widely depending on the configuration of the NN, a large NN decoder does not necessarily achieve high performance.
We chose the NN configuration with the best balance between the number of total parameters and performance as the base model.

\begin{figure*}[tb]
    \centering
    \includegraphics[width=\linewidth]{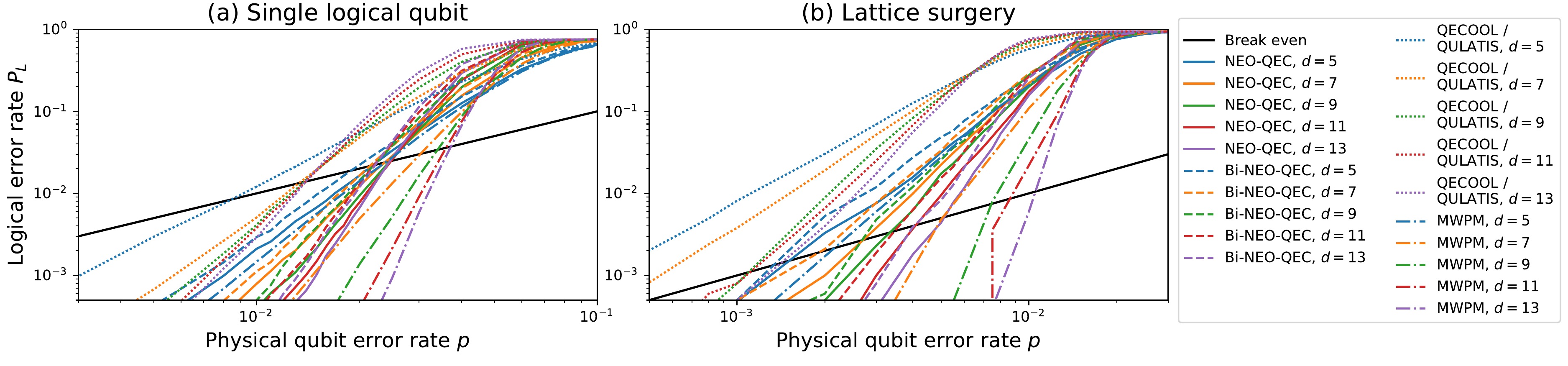}
    \caption{Logical error rate performance of MWPM, QECOOL/QULATIS, and NEO-QEC (FP32 and XNOR-net-based binarized one) with base model NN decoder for (a) single logical qubit protection and (b) the merge-and-split operation of two logical qubits. 
    }
    \label{fig:single_LS_plots}
\end{figure*}

Figure~\ref{fig:single_LS_plots}~(a) shows the results of NEO-QEC with the baseline NN of Tab.~\ref{tab:BNN_model_search}~(solid lines), QECOOL (dotted lines), and MWPM (dash-dot lines).
The performance of the NEO-QEC is still low compared with MWPM, which is a software-based decoder with a batch-processing manner.
The NEO-QEC performance is higher than that of the original QECOOL for any code distance $d$. The pseudo threshold values, which represent intersections of each plot and break-even line, improve by 70\% compared with those of the QECOOL. In addition, the threshold value of NEO-QEC is $p=2.5\%$, whereas those of MWPM and QECOOL are 4.5\% and 1.5\%, respectively.
%This problem could be solved by adjusting the structure of the NNs and changing the training method.

Next, we show the performance of the NEO-QEC algorithm for merge-and-split operations on two logical qubits described in Fig.~\ref{fig:lattice_surgery_example}. 
Figure~\ref{fig:single_LS_plots}~(b) shows the results of NEO-QEC with the baseline NN for the LS procedure. The NEO-QEC performance is higher than that of the QULATIS decoder for any code distance $d$, and the pseudo threshold values improve by 3 to 13 times compared with QULATIS. The threshold values of NEO-QEC, QULATIS, and MWPM are 1.0\%, 0.6\%, and 2.0\%, respectively.

\subfile{./table_BNN_model_search}

\subsection{Binarized NN decoder}

In this subsection, we demonstrate the performance of the NEO-QEC with a binarized NN. The QEC workload is the same as that in the previous subsection. We evaluate the error correction performance of NEO-QEC when XNOR-net-based binarization\cite{Rastegari2016XNORNetIC} is used. Figure~\ref{fig:single_LS_plots}~(a) compares the performances of the NEO-QEC with FP32 and XNOR-net-based binarized NNs (solid and dashed lines, respectively) and QECOOL (dotted lines) for the idling operations of a single logical qubit. The base model NN decoder of Tab.~\ref{tab:BNN_model_search} is used for both FP32 NEO-QEC and binarized NN. The performance of the NEO-QEC with the binarization techniques is better than that of the original QECOOL for any code distance $d$, and its degradation is moderate compared with FP32. The threshold value of NEO-QEC with binarized NN is $p=2.3\%$, whereas that of FP32 NEO-QEC is 2.5\%.

Figure~\ref{fig:single_LS_plots}~(b) shows the result of the binarized NEO-QEC and QULATIS for LS. The results have the same trend as in the case of single logical qubit protection; the threshold values of NEO-QEC with FP32 and binarized NNs are approximately equal at 1.0\%, and the pseudo threshold values are slightly degraded due to binarization.

\subsection{Estimation of computational cost for NN-based decoder\label{subsec:computational_cost}}
In this subsection, we estimate the computational cost of the NN decoder for the case of a single qubit operations. The number of multiplications per convolutional layer $\text{M}_{\text{conv}}$ is calculated as follows:
\begin{align*}
\text{M}_{\text{conv}} =&(\text{kernel size})^2 \times (\text{input channels}) \\&\times (\text{output channels}) \times (\text{output size})^2.
\end{align*}
Then, we estimate the number of multiplications required for one inference of the base NN model in the case of code distance $d$ as follows.
\begin{align*}
(\text{1st layer}): \quad& 7^2 \times 10 \times 16 \times (2d-1)^2 =\quad 7840(2d-1)^2,\\
(\text{2nd layer}): \quad& 5^2 \times 16 \times 16 \times (2d-1)^2 =\quad 6400(2d-1)^2,\\
(\text{3rd layer}): \quad& 5^2 \times 16 \times 16 \times (2d-1)^2 =\quad 1600(2d-1)^2,\\
(\text{Total}):     \quad& (7840+6400+1600)(2d-1)^2             \\&=\quad 15840(2d-1)^2.
\end{align*}
Therefore, sufficient number of NPUs are required to perform $15840(2d-1)^2$ multiplications within 1~$\mu$s for online decoding. 

Here, we assume an NN decoder architecture consisting of $(2d-1)^2$ NPUs and estimate the power consumption of the decoder. Because the process of a convolutional layer can be easily parallelized, the required throughput for each NPU is 15840 multiplications per 1~$\mu$s with the architecture. Hence we assume a 16~GHz clock frequency of NPU that meets the required throughput.

To achieve a lower power decoder, we use ERSFQ logic to design our decoder\cite{kirichenko2011zero}. As we explained in Section~\ref{subsec:SFQ_intro}, we can estimate power consumption of the ERSFQ circuit with Eq.~\ref{eq:ERSFQ}. 
Therefore, the power consumption of the NPU in a 4-K environment $P_{\text{NPU}}$ is as follows:
\begin{align}\nonumber
    P_{\text{NPU}} &= 11.2_{[\text{mA}]} \times 16_{[\text{GHz}]} \times (2.068 \times 10^{-15})_{[\text{Wb}]} \times 2 
    \\&= 0.742_{[\text{$\mu$W}]}.
\end{align}

We assume the idling operations of a distance-9 single logical qubit and the operating frequency of NPUs and QECOOL units are 16 and 2~GHz, respectively. 

The power consumption of the NN decoder per logical qubit $P_{\text{NN}}$ is
\begin{align}\nonumber 
    P_{\text{NN}} &= 0.742_{\text{[$\mu$W/NPU]}} \times (2\times9-1)^2_{\text{[NPUs/logical qubit]}} \\
    &= 214.6_{\text{[$\mu$W/logical qubit]}}.
\end{align}
The power consumption of QECOOL units per logical qubit $P_{QECOOL}$ is 400.3~$\mu$W from the results of Ref.~\cite{ueno2021qecool}.
Therefore, the total power consumption of the NEO-QEC decoder per logical qubit $P_{\text{decoder}}$ is as follows:
\begin{align}\nonumber 
    P_{\text{decoder}} &= P_{\text{NN}} + P_{\text{QECOOL}}\\
    &= 214.6 + 400.3 = 614.9_{[\text{$\mu$W/logical qubit}]}.
\end{align}
The power budget of the 4-K temperature region of dilution refrigerators is supposed to be 1~W\cite{cryogenic_quantum}. 
Thus, we expect that 
\begin{align*}
1_{\text{[W]}} / 614.9_{[\text{$\mu$W/logical qubit}]} \approx \textbf{1626}~\text{logical qubits}
\end{align*}
can be protected in a cryogenic environment with our NEO-QEC architecture.

%% file: NEOQECchap/table_BNN_model_search.tex
\newcommand{\Bf}{\textbf}
\begin{table*}[tb]
\centering
\caption{Total parameter sizes and pseudo thresholds of NEO-QEC (FP32) for various NN configurations.\label{tab:BNN_model_search}}
\scalebox{0.85}{
\begin{tabular}{|c|c|c|c|c||r|c|c|c|c|c||c|c|c|c|c|} \hline
\multirow{3}{*}{K} & \multirow{3}{*}{\begin{tabular}[c]{@{}l@{}}\# of \\ layers\end{tabular}} & \multirow{3}{*}{\begin{tabular}[c]{@{}l@{}}Kernel\\ sizes\end{tabular}} & \multirow{3}{*}{\begin{tabular}[c]{@{}l@{}}Hidden\\ channels\end{tabular}} & \multirow{3}{*}{Note} & \multirow{3}{*}{\begin{tabular}[c]{@{}l@{}}\# of total\\ parameters\end{tabular}} & \multicolumn{10}{c|}{Pseudo threshold $(\times 10^{-2})$} \\ \cline{7-16}
                   &                    &                                                       &    &                   &              & \multicolumn{5}{c||}{Single logical qubit} &\multicolumn{5}{c|}{Marge-and-split operation (LS)} \\ \cline{7-16}
                   &                    &                                                       &    &                   &              &  $d=5$   &   $d=7$   &   $d=9$   &   $d=11$  &   $d=13$  &   $d=5$   &   $d=7$   &   $d=9$   &   $d=11$  &  $d=13$    \\ \hline \hline
  \multicolumn{4}{|c|}{N/A}                                                                          & QECOOL/QULATIS    & N/A          &  0.837   &   1.327   &   1.375   &   1.393   &   1.393   &   0.012   &   0.019   &   0.103   &   0.104   &   0.120    \\ \hline \hline
\multirow{8}{*}{3} & \multirow{4}{*}{2} & \multirow{2}{*}{$7\times7$, $5\times5$}               & 9  & Most lightweight  &  $4,428$     &  2.100   &   2.306   &   2.363   &   2.374   &   2.363   &   0.018   &   0.091   &   0.150   &   0.150   &   0.232    \\ \cline{4-16}
                   &                    &                                                       & 16 &                   &  $7,872$     &  2.151   &   2.349   &   2.436   &   2.418   &   2.413   &   0.118   &   0.216   &   0.288   &   0.371   &   0.461    \\ \cline{3-16}
                   &                    & \multirow{2}{*}{$7\times7$, $7\times7$}               & 9  &                   &  $5,292$     &  2.127   &   2.328   &   2.408   &   2.398   &   2.387   &   0.115   &   0.211   &   0.285   &   0.350   &   0.438    \\ \cline{4-16}
                   &                    &                                                       & 16 &                   &  $9,408$     &  2.152   &   2.360   &   2.448   &   2.434   &   2.424   &   0.112   &   0.218   &   0.309   &   0.365   &   0.462    \\ \cline{2-16}
                   & \multirow{4}{*}{3} & \multirow{2}{*}{$7\times7$, $5\times5$, $5\times5$ }  & 9  &                   &  $6,453$     &  2.178   &   2.365   &   2.450   &   2.479   &   2.447   &   0.088   &   0.175   &   0.255   &   0.314   &   0.357    \\ \cline{4-16}
                   &                    &                                                       & 16 &                   &  $14,272$    &  2.172   &   2.378   &   2.476   &   2.480   &   2.482   &   0.108   &   0.164   &   0.281   &   0.347   &   0.413    \\ \cline{3-16}
                   &                    & \multirow{2}{*}{$7\times7$, $7\times7$, $7\times7$}   & 9  &                   &  $9,261$     &  2.146   &   2.367   &   2.440   &   2.461   &   2.433   &   0.123   &   0.215   &   0.307   &   0.381   &   0.464    \\ \cline{4-16}
                   &                    &                                                       & 16 &                   &  $21,952$    &\Bf{2.208}&   2.397   &   2.502   & \Bf{2.502}&\Bf{2.509} &   0.117   &   0.219   &   0.314   &   0.395   &   0.489    \\ \hline
\multirow{2}{*}{4} & 2                  & $7\times7$, $5\times5$                                & 16 &                   &  $9,440$     &  2.166   &   2.352   &   2.430   &   2.431   &   2.424   &   0.122   &   0.232   &   0.325   &   0.364   &   0.486    \\ \cline{2-16}
                   & 3                  & $7\times7$, $5\times5$, $5\times5$                    & 16 &  Base model       &  $15,840$    &  2.197   &   2.377   &   2.483   &   2.498   &   2.471   & \Bf{0.129}& \Bf{0.256}& \Bf{0.324}& \Bf{0.406}& \Bf{0.511} \\ \cline{1-16}
\multirow{2}{*}{5} & 2                  & $7\times7$, $5\times5$                                & 16 &                   &  $11,008$    &  2.131   &   2.356   &   2.419   &   2.430   &   2.407   &   0.117   &   0.216   &   0.288   &   0.361   &   0.450    \\ \cline{2-16}
                   & 3                  & $7\times7$, $5\times5$, $5\times5$                    & 16 &                   &  $17,408$    &\Bf{2.208}& \Bf{2.406}& \Bf{2.511}&   2.488   &   2.484   &   0.117   &   0.224   &   0.321   &   0.377   &   0.461    \\ \hline%\cline{3-10}
\end{tabular}
}
\end{table*}

%% file: NEOQECchap/06conclusion.tex
\section{Summary and discussion}
\label{sec:BNN_conclusion}
%%% summary
In this paper, we proposed a NN-based online-QEC algorithm named NEO-QEC for decoding SC and LS with measurement errors by combining a convolutional NN and an existing online greedy decoder. 
We evaluated the decoder performance for a single logical qubit idling operation and an LS procedure with code distances up to 13. 
The decoder demonstrated logical error thresholds of about 2.3\% and 1.3\% for a single logical qubit and LS, respectively. The binarized NN technique enables us to implement the lightweight NN decoder with moderate accuracy degradation. Our decoder performance is comparable to the MWPM decoder. 
Furthermore, we designed an SFQ logic-based NPU for binarized NNs with XNOR operations. We evaluated the circuit characteristics of our architecture supporting NEO-QEC and the error-correcting performance of the algorithm. NEO-QEC has low latency and sufficient power efficiency to operate online decoding in a cryogenic environment. 

%%%future work
In this work, we aimed to design a hardware-efficient NN-based decoder rather than an accurate one. Hence we used a simple CNN with a reasonable number of parameters. The maximum operating frequency (approximately 70 GHz) of the designed NPU indicates its potential to construct more accurate and low-latency NN-based decoders with a large number of parameters or with more complex NN techniques, such as 3-D convolutional or convolutional LSTM layers. 

Although we assume a depolarizing noise model, where errors occur on each data and ancillary qubit independently in each QEC cycle, experimental results of real quantum devices have shown that the relaxation and dephasing times of qubits are different, which has a significant impact on the QEC performance with SCs\cite{huang2019performance}. NN-based decoders have the potential to handle non-uniform errors on qubits by learning a bias of qubits on the real device, either explicitly or implicitly. Our future work will extend our lightweight NN-based decoder to handle a realistic biased noise on real quantum devices.

%%% conclusion
Overall, we have designed a fast and ultra-low-power decoder for SC and LS by taking advantage of the binarized NN technique, and it has been shown to have the potential to operate in a cryogenic environment. It has achieved an essential milestone toward practical FTQC and motivates further research on NN-based SC decoders.